\documentclass[aps,prb,twocolumn,floatfix,preprintnumbers,amsmath,amssymb,superscriptaddress]{revtex4}

\usepackage{graphicx}
\usepackage{dcolumn}
\usepackage{bm}
\usepackage{graphics}
\usepackage{lscape}
\usepackage{color}
\usepackage[dvips]{epsfig}
\usepackage{float}

\begin{document}

\preprint{APS/123-QED}

\title{Probing the interface of Fe$_{3}$O$_{4}$/GaAs thin films by hard x-ray photoelectron spectroscopy}

\author{M. Paul}
\email{paul@physik.uni-wuerzburg.de}
\author{A. M\"uller}
\author{A. Ruff}
\author{B. Schmid}
\author{G. Berner}
\affiliation{Universit\"at W\"urzburg, Experimentelle Physik 4, D-97074 W\"urzburg, Germany}

\author{M. Mertin}
\affiliation{BESSY GmbH, Albert-Einstein-Strasse 15, D-12489 Berlin, Germany}

\author{M. Sing}
\author{R. Claessen}
\affiliation{Universit\"at W\"urzburg, Experimentelle Physik 4, D-97074 W\"urzburg, Germany}
        
\date{\today}

\begin{abstract}
Magnetite (Fe$_3$O$_4$) thin films on GaAs have been studied with HArd X-ray PhotoElectron Spectroscopy (HAXPES) and low-energy electron diffraction. Films prepared under different growth conditions are compared with respect to stoichiometry, oxidation, and chemical nature. Employing the considerably enhanced probing depth of HAXPES as compared to conventional x-ray photoelectron spectroscopy (XPS) allows us to investigate the chemical state of the film-substrate interfaces. The degree of oxidation and intermixing at the interface are dependent on the applied growth conditions; in particular, we found that metallic Fe, As$_{2}$O$_{3}$, and Ga$_{2}$O$_{3}$ exist at the interface. These interface phases might be detrimental for spin injection from magnetite into GaAs.
\end{abstract}



\maketitle

\begin{figure}[htpb]
      \includegraphics[width=0.4\textwidth]{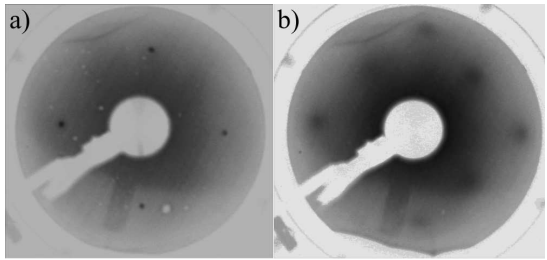}
        \caption[sample2LEED]{\small{a) typical LEED pattern of a GaAs substrate showing the $1\times1$ surface unit cell; E = 35.9\,eV b) LEED pattern of a Fe$_{3}$O$_{4}$ thin film (sample 1) showing the $1\times1$ surface unit cell corresponding to the O sublattice; E = 89.0\,eV}}
        \label{Fig1}
\end{figure}

\begin{figure*}[htbp]  
 \centering
    \includegraphics[width=1\textwidth]{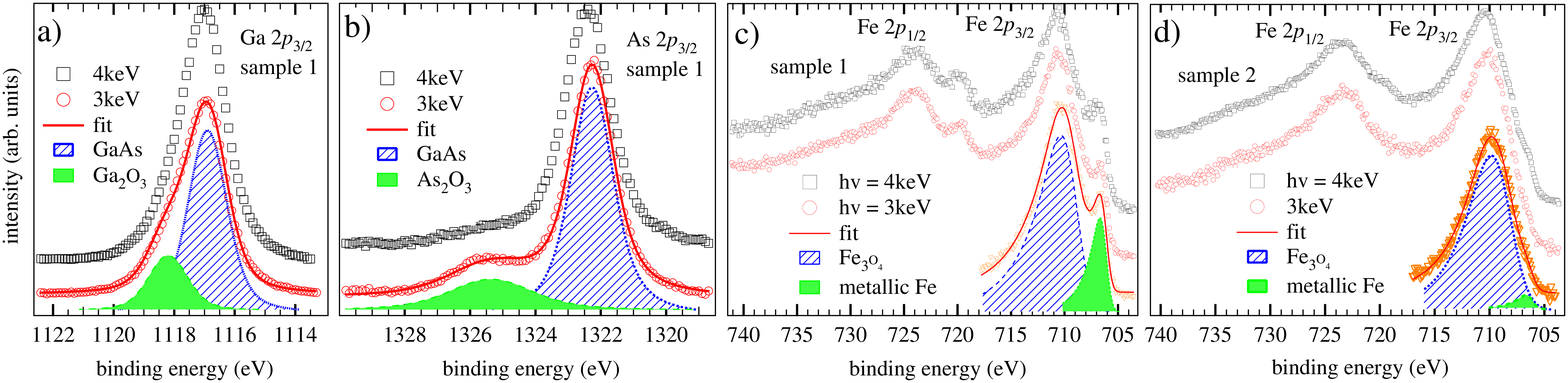}
        \caption[Ga2p level]{\small{HAXPES core level spectra of Fe$_{3}$O$_{4}$/GaAs samples taken with $h\nu$ = 3\,keV and 4\,keV; also shown are the decomposition of the spectra in various components according to a fitting procedure and the resulting fit curves; curves are shifted in vertical direction for clarity a) sample 1: Ga\,2\emph{p}$_{3/2}$ b) sample 1: As\,2\emph{p}$_{3/2}$ c) sample 1: Fe\,2\emph{p} d) sample 2: Fe\,2\emph{p}}}
        \label{Fig2}
 \end{figure*}

The ferrimagnetic iron oxide magnetite (Fe$_3$O$_4$) is ranked among the most attractive materials for  the currently developing field of spintronics.\cite{Hag04} The basic concept of spintronics consists in the design of integrated circuits which use  the electron \textit{spin} for data storage \textit{and} processing of information.\cite{Wol01} In spintronic devices semiconducting materials, used in conventional charge-based electronic chips, and ferromagnetic materials, employed in storage devices, are combined in a new way. Therefore a key element is the integration of magnetic materials like metallic ferromagnets, diluted magnetic semiconductors, or oxidic half-metallic ferromagnets with substrates used in existing semiconductor technology. Magnetite stands out from other feasible magnetic materials due to the following bulk material characteristics: a very high Curie temperature of 858\,K, a predicted spin-polarization of 100\,\% at the Fermi level,\cite{Yan99} and a conductivity of $2.5\cdot10^{4}\,(\Omega m)^{-1}$ at room temperature\cite{Mil57} which matches quite well the value of semiconducting materials. For a film/substrate structure without buffer layer the latter two features are crucial to facilitate spin-injection into the semiconducting host via an ohmic contact.\cite{Sch00}
\\Hence there exists a clear need to study the growth behavior and thin film properties of magnetite on semiconducting substrates like GaAs. Moreover, a detailed knowledge of the actual interface structure and stoichiometry is desirable in order to correlate it with the magnetic and spin transport properties. Interface issues as the occurrence of mixed phases could prevent a successful growth or at least influence material properties in an undesired way, e.g., limit the degree of spin-polarization at near-interface layers or surfaces.
\\Previous work done on thin film growth of Fe$_3$O$_4$ mostly utilized oxide substrates. Recent interest has been directed at the use of semiconducting substrates.\cite{Lu04, Fer07, Jai05} However, there are only sparse experimental reports\cite{Pre03,Jai05} on interface chemistry and interface reactions, although based on thermodynamic considerations such interfaces might not be stable.\cite{Hub96}
\\In this work, we examine this problem and investigate the chemical nature of Fe$_3$O$_4$ films grown on GaAs(100) substrates and their respective interfaces. Hard x-ray photoelectron spectroscopy (HAXPES) is an ideal tool to study the electronic structure and chemical state of these films. In particular, the larger information depth due to photon energies in the hard x-ray regime allows to probe the interface between film and substrate in a non-destructive way. Changing the photon energy while looking at the same core level permits for depth profiling of the sample with respect to the specific element or chemical species under consideration. To give numbers, the ineleastic mean free path increases from 38\,{\AA} to 51\,{\AA}, from 32\,{\AA} to 46\,{\AA} and from 30\,{\AA} to 43\,{\AA} for the Fe\,2\emph{p}$_{3/2}$, Ga\,2\emph{p}$_{3/2}$,  and As\,2\emph{p}$_{3/2}$ core level, respectively, upon changing the photon energy from 3\,keV to 4\,keV.\cite{Tan05}
\\Fe$_{3}$O$_{4}$/GaAs(100) samples were grown in a ultra-high vacuum chamber equipped with an electron-beam evaporator with built-in flux meter, a gas inlet system for oxygen of ultra-high purity, and a LEED (low energy electron diffraction) optics for surface monitoring. Beforehand GaAs substrates were cut from a Si-(n-)doped wafer, etched with highly concentrated sulphuric acid, and rinsed with deionized water, both under flowing conditions. As an \textit{in situ} treatment sample 1 was sputtered  (Ar$^{+}$, energy 1\,keV) and annealed to 820\,K, while sample 2 was annealed only  to 770\,K prior to actual film growth. As we verified by XPS measurements these substrate treatments result in clean GaAs without its native oxides. The following growth conditions were applied: Fe film growth at room temperature and post-oxidation at 700\,K and p(O$_{2})= 1\cdot10^{-6}$\,mbar for 30\,min (sample 1) and at 600\,K and p(O$_{2})= 5\cdot10^{-5}$\,mbar for 3\,min (sample 2). The deposited Fe film thickness was 23\,{\AA} (sample1) and 36\,{\AA} (sample 2), respectively, as derived from the flux meter calibrated against a quartz crystal monitor.
\\A typical LEED pattern of a cleaned GaAs(100) substrate, which exhibits the $1\times1$ surface unit cell, is displayed in Fig.~\ref{Fig1}a. On such substrate surfaces Fe, which is known to grow epitaxially on GaAs(100),\cite{Xu98} was deposited at room temperature. As is known from the literature the growth mode depends on the substrate reconstruction and temperature, and is, e.g., three-dimensional on Ga-rich GaAs(100)-c($8\times2$) with islands coalescing above 4 monolayers\cite{Cha86} or predominantly layer-by-layer on As-rich GaAs(100)-($2\times4$).\cite{Kne96} The epitaxial growth of Fe in our case was confirmed by good quality LEED patterns (not shown here). 
\\LEED images of Fe films post-oxidized to Fe$_{3}$O$_{4}$ (see Fig.~\ref{Fig2}b) show a square unit cell with a lattice constant of about 3\,{\AA}. Additional spots were not present. Naively, one would expect a lattice constant of 5.9\,{\AA} for magnetite (corresponding to a $2\times2$ unit cell with respect to the observed one) or a ($\sqrt{2}\times\sqrt{2}$)R45$^{\circ}$ superstructure, giving rise to a lattice constant of 8.4\,{\AA}. The latter is typically observed for magnetite single crystals or thicker films. Since the ($\sqrt{2}\times\sqrt{2}$)R45$^{\circ}$ superstructure is due to the polar nature of the Fe$_{3}$O$_{4}$ stacking sequence, its absence can be related to the very small film thickness. We are thus led to interpret the observed $1\times1$ unit cell with a lattice constant of roughly 3\,{\AA} as signature of the oxygen sublattice, which is common to all Fe oxides (FeO, Fe$_{3}$O$_{4}$, Fe$_{2}$O$_{3}$). The main reason for the missing $2\times2$ and the observed broad diffraction spots might be disorder due to amorphous interface phases.

HAXPES experiments were carried through at room temperature and without further surface treatment to avoid a change in chemical composition using beamline KMC-1 at BESSY in Berlin. The total energy resolution was 0.74\,eV and 0.90\,eV at a photon energy of 3\,keV and 4\,keV, respectively, as was checked by measuring the Au\,4\emph{f}$_{7/2}$ core level with an intrinsic linewidth of 0.25\,eV (full width half maximum). The shown spectra have been shifted to correct for charging by setting the O\,1\emph{s} binding energy to 530.1\,eV which is known to be the same in all Fe oxides.\cite{Gota01} 
\\In Fig.~\ref{Fig2} core level spectra measured with photon energies of 3\,keV and 4\,keV on sample 1 and sample 2 are shown. The Ga\,2\emph{p}$_{3/2}$ level (see Fig.~\ref{Fig2}a) is composed of a main component due to GaAs at 1116.9$\pm$0.2\,eV binding energy, which matches perfectly the literature value for GaAs\cite{Mas85}, and a smaller oxide peak at 1.3\,eV higher binding energy, which appears in the 3\,keV spectrum as small shoulder and is hardly visible in the spectrum taken with 4\,keV photons. Due to its chemical shift with respect to the main line the oxide peak can be attributed to Ga$_{2}$O$_{3}$.\cite{Mas85} The main component due to GaAs is enhanced in the 4\,keV spectrum with higher information depth while the oxide component is stronger in the more surface and interface sensitive spectrum. 
\\Similar conclusions can be drawn in the case of the As\,2\emph{p}$_{3/2}$ core level shown in Fig.~\ref{Fig2}b. Here the main component has a binding energy of 1322.4$\pm$0.1\,eV, which is comparable to the literature\cite{Mas85}, where values of 1322.8\,eV for GaAs and 1322.4\,eV for elemental As are reported. Taking into account the relevant photoionization cross sections, asymmetry parameters, inelastic electron mean free paths, and the analyzer transmission function, the intensity ratios of the Ga\,2\emph{p}$_{3/2}$ to the As\,2\emph{p}$_{3/2}$ main component range from 0.82 to 0.98 for different samples and photon energies and are therefore close to the ideal stoichiometry for GaAs. On this account we assign the As\,2\emph{p}$_{3/2}$ main component to GaAs and not to elemental As since there is no indication for a significant off-stoichiometric amount of excess As with respect to Ga. After having identified the As\,2\emph{p}$_{3/2}$ main component,an additional component or shoulder is indeed not seen at a chemical shift of 0.6\,eV\cite{Hol94,Liu03} higher binding energy, which then would indicate elemental As. The oxide As\,2\emph{p}$_{3/2}$ component appears at 3.1\,eV higher binding energies and hence can be attributed to As$_{2}$O$_{3}$.\cite{Mas85} The main component due to GaAs is enhanced in the 4\,keV spectrum with higher information depth while the oxide component is stronger in the more surface and interface sensitive spectrum.
\\As has been mentioned above the O\,1\emph{s} main peak (not shown here) serves as energy reference with a binding energy of 530.1\,eV. We also find a feature at 1.6\,eV higher binding energy which presumably stems from OH groups at the surface\cite{Liu98} since it decreases upon changing the photon energy from 3\,keV to 4\,keV. This contribution obviously reflects the amount of contamination due to exposure to air and is not related to sample preparation conditions.
\\In Fig.~\ref{Fig2}c and d the Fe\,2\emph{p} spectra of sample 1 and sample 2  are depicted. Concentrating on the more intense Fe\,2\emph{p}$_{3/2}$ part of the spectrum, for sample 1 (see Fig.~\ref{Fig2}c) one can clearly distinguish two spectral features, the main component at 710.1$\pm$0.3\,eV and a smaller peak at the lower binding energy side of the main line, shifted by 3.6\,eV. The main component is readily attributed to Fe$_{3}$O$_{4}$, while the second peak is due to metallic Fe. The same two features can be discerned in the Fe\,2\emph{p} spectrum of sample 2 (Fig.~\ref{Fig2}d) although the metallic Fe related feature appears here only as a weak shoulder. For both samples the metallic component is stronger in the spectrum taken at higher photon energy, i.e., with larger information depth. Since the amount of metallic Fe in sample 2 was too small to be detected by conventional XPS using monochromated Al-K$_{\alpha}$ radiation (1486.6\,eV) we conclude that the metallic Fe is not located at the surface but deeper in the bulk, probably at or near to the interface. The spectrum taken at $h\nu$\,=\,3\,keV on sample 2 exhibits only a marginal contribution of metallic Fe and hence displays most clearly the overall spectral shape of the mixed valence state of Fe$_{3}$O$_{4}$:\cite{Gota01} the  2\emph{p}$_{3/2}$ peak lies roughly at 710.5\,eV and the structure between the spin-orbit split peaks is smeared out. In case of Fe$_{2}$O$_{3}$ the Fe$^{3+}$ charge-transfer satellite should be visible at  719\,eV, and for FeO the Fe$^{2+}$ satellite should appear at 715.5\,eV. Both signatures are not seen here.
\\To quantify our results, for each element its relative amount in a certain chemical species as derived from numerical fits (see Fig.\,2) is summarized in Table \ref{tab:Tab1}. The values support the statement that the oxides Ga$_{2}$O$_{3}$ and As$_{2}$O$_{3}$ are located at the interface or at least nearer to the surface, while the metallic Fe is situated deeper in the bulk or at the interface. The OH-groups are adsorbed at the surface.
\\Assuming a layered structure of the samples, we can calculate the thicknesses of the As and Ga oxide layers at the interface using the equation in footnote \footnote[38]{$d_{ov}=\lambda_{sub} \cos \cdot \theta \cdot \ln (\frac{D_{sub}I_{ov}}{D_{ov}I_{sub}}+1)$, where $\lambda_{sub}$ is the electron mean free path, $\theta$ the electron emission angle from the surface normal, $I_{ov}$ and $I_{sub}$ the XPS peak intensities of overlayer (oxide) and substrate level and $D_{ov}$ and $D_{sub}$ the atomic densities of the element in overlayer and substrate in $mol/cm^3$} (see also Ref. \onlinecite{Liu03}) from the ratios in Table \ref{tab:Tab1}. The As$_{2}$O$_{3}$ layer thickness amounts to 4.9$\pm$1.6\,{\AA} and 1$\pm$1\,{\AA}, the Ga$_{2}$O$_{3}$ layer thickness to 4.0$\pm$1.0\,{\AA} and 1.2$\pm$0.7\,{\AA} for sample 1 and 2, respectively. The thicknesses for both investigated samples differ substantially and are only about one monolayer for sample 2. We finally arrive at a dimensioned sketch of the vertical structures and compositions of samples 1 and 2 displayed in Fig.~3.

\begin{table*}[htpb]
\caption{\small{XPS signal of different chemical species normalized to the total XPS signal for that element}}
    \label{tab:Tab1}
\begin{ruledtabular}
\begin{tabular}{lcccc}
        species  & sample 1, 3\,keV & sample 1, 4\,keV & sample 2, 3\,keV  & sample 2, 4\,keV \\
        \hline
        As: As$_{2}$O$_{3}$/(As$_{2}$O$_{3}$ + GaAs) & 0.21 & 0.08 & 0.07  & 0   \\
        Ga: Ga$_{2}$O$_{3}$/(Ga$_{2}$O$_{3}$ + GaAs) & 0.23 & 0.11 & 0.10 & 0.02 \\
        Fe: Fe/(Fe + Fe$_{3}$O$_{4}$) & 0.19 & 0.23 & 0.04 & 0.07 \\
        O: OH (surface)/(OH + Fe$_{3}$O$_{4}$) & 0.79 & 0.77 & 0.57 & 0.32\\

\end{tabular}
\end{ruledtabular} 
\end{table*}

\begin{figure}[htbp]        
        \centering
    \includegraphics[width=0.4\textwidth]{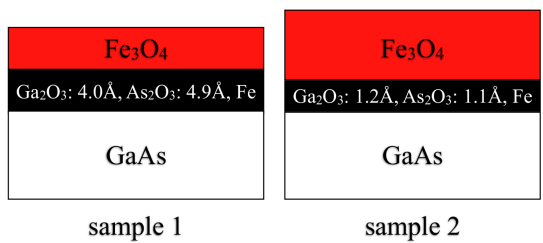}
        \caption[film structure]{\small{simplified sketch of the proposed vertical sample structures with phases and layer thicknesses as indicated}}
        \label{Fig3}
\end{figure} 

The presence of metallic Fe at or near the interface can hardly be attributed to unsufficient oxidizing conditions at the near-interface region since with an applied O$_{2}$ exposure of 1800\,L for both GaAs samples the oxidation thickness of Fe should exceed 68\,{\AA} (value for 1500\,L) \cite{Kim00b} and therefore be enough to fully oxidize the Fe film. Instead, the present results can be conclusively explained by an interface reaction of Fe$_{3}$O$_{4}$ and GaAs to metallic Fe, As$_{2}$O$_{3}$, and Ga$_{2}$O$_{3}$.
\\To better understand the results at hand concerning As oxides and Ga oxides we recall some experimental findings on the oxidation of GaAs. Experiments on the thermal oxidation of GaAs have shown that at high temperatures (800-1000\,K) primarily polycrystalline Ga$_{2}$O$_{3}$ and possibly GaAsO$_{4}$ are formed.\cite{Wil81,Hol94} At lower oxidation temperatures the resulting products are amorphous and again mainly Ga$_{2}$O$_{3}$ and a smaller fraction of elemental As at the oxide/GaAs interface. As$_{2}$O$_{3}$ is additionally found for oxidation with molecular oxygen at low and intermediate temperatures. Transferring these results to our case the oxidation conditions for the GaAs substrate in Fe$_{3}$O$_{4}$/GaAs apparently are weak because of rather low temperatures, small oxygen partial pressure, and the presence of the easily oxidized Fe. In accordance, no GaAsO$_{4}$ has been observed which could easily be detected by XPS due to the large chemical shift of about -4.8\,eV with respect to the GaAs component in both As and Ga core levels. On the other hand, given weak oxidation conditions, elemental As should be found, which we do not. Probably the formation of elemental As is hindered by kinetic factors in our case.
\\Former publications \cite{Lu04,Lu05,Pre03} on Fe$_{3}$O$_{4}$/GaAs samples cover their fabrication by either MBE as in this study or pulsed laser deposition (PLD). Lu \textit{et al.}\cite{Lu04} found that Fe$_{3}$O$_{4}$ grows (100) oriented on GaAs(100) when post-oxidizing a Fe film grown by MBE. Interestingly, they have measured a Fe\,2\emph{p} spectrum by conventional XPS indicating small amounts of metallic Fe, similar to what we find, but this fact was not commented there.\cite{Lu04} However, in a later X-ray magnetic circular dichroism study\cite{Lu05} on similarly prepared samples they did not confirm the presence of metallic Fe and reported instead that above a critical Fe$_{3}$O$_{4}$ film thickness of 3\,nm a FeO interface layer forms and increases with film thickness. The presence of interfacial FeO was attributed to oxygen defects. We note that the presence of interfacial FeO is explained with an interface reaction between Fe$_{3}$O$_{4}$ and GaAs producing FeO and Ga and As oxides, similar to our case.
\\Preisler \textit{et al.}\cite{Pre03} proposed that their obtained (111) oriented polycrystalline growth of Fe$_{3}$O$_{4}$ on GaAs(100) by means of PLD is triggered by the presence of an amorphous interface. Moreover, they saw a strong shoulder in the XPS Ga\,3\emph{d} spectrum measured with Al-\textit{K}$_{\alpha}$ radiation indicating Ga-Fe bonding. We clearly do not see evidence for a (111) orientation, a polycrystalline film structure, or Ga-Fe bonding in our results. However, the moderate quality of the obtained LEED patterns of our very thin films could be linked to the presence of amorphous Ga and As oxide interface phases. The finding of a Ga-Fe species by Preisler \textit{et al.} can be explained by a stronger intermixing and a reduction of Fe$_{3}$O$_{4}$ to Fe at the interface which in turn could be induced by the use of the PLD technique .
\\With the future application of Fe$_{3}$O$_{4}$/GaAs in spintronics in mind, the influence of interface phases such as Fe, As$_{2}$O$_{3}$, and Ga$_{2}$O$_{3}$ on the magnetic properties of Fe$_{3}$O$_{4}$ films is an important issue which demands for further investigations. It has, e.g., been shown that a Fe$_{3}$O$_{4}$/Fe  bilayer possesses antiparallel magnetic coupling.\cite{Kim00b} Further, a metallic interface layer with high conductivity is expected to decrease the efficiency of spin injection considerably.\cite{Sch00} At last, As$_{2}$O$_{3}$ and Ga$_{2}$O$_{3}$ could act as a spin injection tunneling barrier for the spin-polarized electrons. Concerning the latter point, from our results post-oxidation of the Fe films at lower temperatures seems favorable since it results in considerably smaller thicknesses of the interfacial oxide layer (see Fig.~3).
\\In summary, we have demonstrated the successful growth of Fe$_{3}$O$_{4}$ films on the semiconducting substrate GaAs(100) and have characterized them through depth-resolved investigation of the chemical nature of film  and interface regions. An oxidation of GaAs to As$_{2}$O$_{3}$ and Ga$_{2}$O$_{3}$ near the interface and a simultaneous reduction of Fe$_{3}$O$_{4}$ to Fe has been revealed. For a lower growth temperature of 600\,K the amounts of oxidized GaAs and metallic Fe were less compared with post-oxidation at 700\,K. As has been demonstrated HAXPES is a useful and essential method for the depth-resolved characterization of chemical phases in thin film structures and is superior to conventional XPS for the identification of interface phases.
\\The authors thank A. Gloskovskij, F. Casper, J. Barth, and S. Tabor for their experimental support during beamtimes at BESSY. They acknowledge financial support by BMBF through project 05 KS7WW3. M.P. acknowledges financial support through the scholarship program by DAAD.

\end{document}